\documentstyle[12pt,epsf]{article}
\newcommand{\OO}{{\cal O}}
\newcommand{\CC}{{\cal C}}
\newcommand{\II}{{\cal I}}
\newcommand{\SS}{{\cal S}}

\newcommand{\TT}{{\cal T}}

\newcommand{\QQ}{{\cal Q}}

\newcommand{\wt}{\widetilde}

\newcommand{\uu}{u}

\newcommand{\be}{\begin{equation}}
\newcommand{\ee}{\end{equation}}
\newcommand{\ben}{\begin{eqnarray}\displaystyle}
\newcommand{\een}{\end{eqnarray}}
\newcommand{\refb}[1]{(\ref{#1})}

\newcommand{\sectiono}[1]{\section{#1}\setcounter{equation}{0}}

\begin{document}
{}~
\hfill\vbox{\hbox{hep-th/0012251}\hbox{CTP-MIT-3064}
\hbox{PUPT-1972} \hbox{MRI-P-001201}
}\break

\vskip 1.4cm

\centerline{\large \bf String Field Theory Around the Tachyon Vacuum}

\vspace*{5.0ex}

\centerline{\large \rm Leonardo Rastelli$^a$, Ashoke Sen$^b$ and Barton
Zwiebach$^c$}

\vspace*{6.5ex}

\centerline{\large \it ~$^a$Department of Physics }

\centerline{\large \it Princeton University, Princeton, NJ 08540}

\centerline{E-mail:
        rastelli@feynman.princeton.edu}

\vspace*{2ex}

\centerline{\large \it ~$^b$Harish-Chandra Research
Institute\footnote{Formerly Mehta Research Institute of Mathematics
and Mathematical Physics}}

\centerline{\large \it  Chhatnag Road, Jhusi,
Allahabad 211019, INDIA}

\centerline{E-mail: asen@thwgs.cern.ch, sen@mri.ernet.in}

\vspace*{2ex}

\centerline{\large \it $^c$Center for Theoretical Physics}

\centerline{\large \it
Massachussetts Institute of Technology,}

\centerline{\large \it Cambridge,
MA 02139, USA}

\centerline{E-mail: zwiebach@mitlns.mit.edu}

\vspace*{8.5ex}

\centerline{\bf Abstract}

\bigskip

Assuming that around the tachyon vacuum the
kinetic term of cubic open string
field theory is made purely of ghost
operators we are led to gauge invariant
actions which manifestly implement the absence
of open string dynamics around
this vacuum. We test this proposal by showing
the existence of lump solutions of
arbitrary codimension
in this string field theory.
The key ingredients in this
analysis are certain assumptions about the
analyticity properties of tachyon
Green's functions. With the help of some further
assumptions about the properties
of these Green's functions,
we also calculate the ratios of
tensions of lump solutions of different dimensions.
The result is in perfect
agreement with the known answers for the ratios of
tensions of D-branes of
different dimensions.

\vfill \eject

\baselineskip=16pt

\tableofcontents

\sectiono{Introduction and Summary} \label{s1}

The 26-dimensional bosonic string theory contains D-$p$-branes for all
integer $p\le26$. Each of these D-$p$-branes has a tachyonic mode. It has
been
conjectured \cite{9902105,9904207} that there is a local minimum of the
tachyon potential which describes the closed string vacuum without any
D-brane. At this minimum the negative contribution from the tachyon
potential exactly cancels the tension of the D-brane. Furthermore,
since the brane disappears, this vacuum cannot support
conventional open string excitations.
It has also been
conjectured that a codimension $q$ lump solution on the D-$p$-brane
represents a D-$(p-q)$-brane in the same theory. Support for these
conjectures comes from the analysis of the world-sheet
theory~\cite{9902105,9811237,9402113,9404008,9406125,0003101,0010247},
cubic
open string field
theory~\cite{KS,9912249,0002237,0002117,0003031,
0005036,0006240,0007153,
0008033,0008053,0008101,0008252,0009105,0010190,0011238},
noncommutative limit of the effective field theory of the
tachyon~\cite{0003160,0005006,0005031,0006071,0010060,0008214}, background
independent open string field theory~\cite{0009103,
0009148,0009191, david},
as well as various simple
models of tachyon
condensation~\cite{0003278,0008227,
0008231,0009246,0011226,0012080}.

One of the promising approaches to the study of this
subject matter
involves the use of
cubic open string field theory (SFT) \cite{WITTENBSFT}.
Indeed,
much work
has already been done with it
using the level
truncation scheme \cite{KS,9912249,0002237}.
While complete analytic solutions
of the equations of motion of this string field theory
are still
missing, the solutions
found by combination of analytic and numerical work
gave some of the strongest evidence for the conjectures,
as well as demonstrated that string field theory contains
non-perturbative physics. More recently, work in boundary
string field theory (B-SFT)
\cite{9208027, 9210065,9303067,9303143,9311177,0011033},
$-$ a version of
string field theory incorporating a certain
degree of background independence, $-$
confirmed the form of the tachyon
potentials proposed  in \cite{0008231,0009246}
and established
conclusively the energetics aspect of the tachyon conjectures
\cite{0009103,0009148,0009191}.
Whereas in conventional
SFT the tachyon {\it string field} condenses, namely an infinite
number of scalars acquire expectation values, in B-SFT only the
tachyon field acquires a
vacuum expectation value. This is quite
fortunate, as it appears
to be very difficult to formulate B-SFT
in generality. Due to this problem in B-SFT,
at present cubic string field theory seems to be the only complete
framework where we can analyze concretely the fate of the open
string field excitations around the vacuum where the D-brane
has disappeared. Indeed, it seems to provide
a completely non-singular framework to study the tachyon
vacuum. Starting from a D-brane background one appears
to reach the tachyon vacuum with room to spare and without
any indications of singularities in field variables.

In principle the analysis
of the tachyon vacuum
is straightforward. One must: (i) find
the classical solution $\Phi_0$  representing the vacuum with no
D-brane, (ii) expand the string field action setting $\Phi =
\Phi_0 + \Psi$, where $\Psi$ is the fluctuation field,
and (iii) analyze the spectrum of $\Psi$ using the resulting
kinetic term. Nevertheless, in practice this has not been
simple to carry out, first and foremost because there is no
known closed form expression for $\Phi_0$ yet.  The accurate
numerical approximations to $\Phi_0$ may allow, however, an
analysis of the spectrum around the vacuum \cite{0008033,ellwood}.

In this paper we shall analyze the problem from a different angle. Instead
of trying to construct the classical solution $\Phi_0$,
expand
the SFT action around $\Phi_0$,
and attempt field redefinitions to bring
the kinetic term to a simple form,  we shall
make an inspired guess about the form of the SFT action expanded around the
tachyon vacuum.
Then we will try to check that this action satisfies the
various consistency requirements. Since the string field theory action is
cubic in the string field, shifting the string field by a classical solution
does not change the cubic interaction term. Thus we only need to guess the
quadratic term of the shifted action. The requirement that the new
action is
obtained from the original one by a shift of the string field by a classical
solution of the equations of motion puts constraints on the possible choices
of the quadratic term, $-$ these basically correspond to the requirement that
the shifted action also has an infinite parameter gauge
invariance like the original SFT action. Besides these constraints
the action must satisfy the following additional requirements:
\begin{enumerate}
\item
The kinetic operator must have vanishing cohomology.
This would imply absence
of physical open string states around the tachyon vacuum.

\item
The action must have
classical solutions representing the original D-brane
configuration, as well as lump solutions representing D-branes of all
lower dimensions.
\end{enumerate}

Indeed, our work began with the simple realization that
the reparametrization ghost zero mode $c_0$ was an obvious
replacement for the BRST operator that would lead to vanishing
cohomology, and thus no physical states.\footnote{This possibility
may have occurred to many physicists, but we heard it first from
E. Witten \cite{wittenunp}.}
More striking, however, is the fact that the Riemann surface description
of the star product makes it evident that $c_0$ is a derivation. Thus,
we noted that
replacing the BRST operator $Q_B$  by $\QQ= c_0$ would lead to a string
field
action
\be \label{firsteqn}
\SS (\Psi) \equiv \,-\, {1\over g_0^2}\,\,\bigg[\, {1\over 2} \langle
\,\Psi \,,\, \QQ\, \Psi
\rangle + {1\over 3}\langle \,\Psi \,,\, \Psi *
\Psi \rangle \bigg] \,,  \quad \QQ = c_0 \,,
\ee
that would have the expected (vanishing) spectrum and would
be gauge invariant. It is straightforward to see that more general
operators lead to the same conclusions. In fact
\be
\label{mgen}
\QQ = \sum_{n=0}^\infty a_n \CC_n \,,  \quad \CC_n \equiv c_n + (-1)^n
c_{-n}\,,
\ee
with arbitrary constants $a_n$'s would also lead to gauge
invariant actions with no physical states. As we will show, many of these
actions are related by field redefinitions.

In fact, long time ago, Horowitz {\it et.al.}~\cite{HORO} constructed a
class of formal solutions of the purely cubic string field
theory action~\cite{CUBIC,Hata:1986vq}
which lead to actions of the above form, with an extra regularity
condition  $\QQ|\II\rangle =0$,  where
$|\II\rangle$ is the (formal) identity of the star algebra. In
particular,  these authors noticed that such string field theory would
describe no physical excitations around the vacuum in question.
Furthermore, since a subclass of the above actions is obtained by
shifting the string field in the purely cubic action, which in turn is
obtained by shifting the string field in the original SFT action, we can,
at least formally, regard the above actions
as the result of shifting the string field in the original SFT
action.\footnote{These solutions,
however, have
the unwanted feature that at least formally they have the same energy
density as the original D-brane configuration. This is not what is
expected of the solution describing the tachyon vacuum. This would
suggest that we only consider those $\QQ$'s which {\it do not} satisfy
the regularity condition $\QQ|\II\rangle =0$. In fact, $c_0|\II\rangle
\not= 0$.}

We will not be able to determine here the precise form and
normalization of $\QQ$ and we take the above class of actions
as confimation that the conditions of zero cohomology, gauge
invariance and universality of the kinetic operator can be
satisfied. There may even be more general forms of kinetic
operators built purely from the ghost sector and satisfying these
conditions.
Such kinetic terms make the universality of
the action expanded around the tachyon vacuum manifest. Indeed, since
the tachyon vacuum, representing the closed string vacuum without any
D-brane, should not depend on which D-brane configuration we start from,
we should expect that the string field theory action expanded around this
vacuum should lose all knowledge about the matter conformal field theory
(CFT) describing the original D-brane configuration.\footnote{Since the
interaction term involves correlation function in the combined
matter-ghost CFT, it does depend on the matter
CFT. For this CFT we use the canonical choice corresponding to the
maximal dimension space-filling D-brane. However,
since the three string vertex involves complete overlap of the three
strings the interaction can be made formally background independent by a
rescaling of the string fields.}

Given the class of actions exemplified by eqns. \refb{firsteqn} and
\refb{mgen} the non-trivial part of the problem is to construct the
classical solutions in this string field theory representing the original
D-brane as well as various lump solutions representing the lower
dimensional D-branes, and compare their tensions with the known tensions
of the D-branes. At present we have only been able to partially
determine the space-time
dependence of the solutions and to
calculate the ratio of the tension of a $p$-brane lump to that of a
$(p+1)$-brane lump for arbitrary integer $p$.
Under certain assumptions which will be explained in section \ref{s3},
this computation can be
done without a detailed knowledge of the quadratic term
in the action
as long as it does
not involve any matter sector operator.
The final result for the ratio of tensions involve certain function
$B(\alpha)$ (defined in eqs.\refb{e33}, \refb{es2a}, \refb{es2}) which, if
zero, will produce the desired value. Although we cannot at present give
an analytic proof of the vanishing of this function, we have checked
numerically that this function indeed vanishes to very high degree of
accuracy. This in turn is evidence that our guess about the
SFT action expanded around the tachyon vacuum is the correct one.

The strategy that we follow for this computation is as follows.
Since the kinetic term and hence the propagator does not involve
any matter operator
we are able
to find the exact
momentum dependence of an arbitrary
$n$-tachyon Green's function. From the generating functional $W[J]$
of the
tachyon Green's functions we can find the tachyon effective action by
Legendre transformation.
Missing
the overall ($n$-dependent) normalization of the $n$-tachyon Green's
function,
we only have a partial
knowledge of the effective action. We now assume that the tachyon
equations of motion derived from this effective action
have
a space-time
independent classical solution which can be identified with the original
D-brane. Using this assumption, together with certain other assumptions
about the analyticity of the generating functional of the tachyon Green's
function stated in section \ref{s3}, we show that the tachyon equation of
motion also has lump solution of every codimension.
Some further
assumptions allow us to calculate the ratios of tensions of these lumps
without knowing the detailed form of the effective potential. As already
stated, the result is in perfect agreement with the known answers for the
ratios of tensions of D-branes.

\medskip
Clearly many things remain to be done. First and  foremost is to
reproduce correctly the overall normalization of the D-brane tension. For
this we need to fix unambiguously the form of the kinetic term $\QQ$ in
the shifted vacuum.\footnote{While $\QQ=0$, leading to the
purely cubic action \cite{CUBIC}, is  universal, the resulting background
is not suitable for the present analysis since with
vanishing kinetic term the Feynman rules used for computing the tachyon
effective action are not well defined. In addition, as opposed
to the string field tachyon condensate, the
%bz the singular shift of the
string field taking the original
D-brane configuration to the one with vanishing kinetic
term is annihilated by the BRST operator. Thus the purely cubic
action is unlikely to represent the tachyon vacuum.}
Some aspects of our analyticity assumptions may require
more information about $\QQ$.
The action around the tachyon vacuum is expected to possess
many novel properties. In particular many of the symmetries which are
spontaneously broken in the background of a D-brane are expected to be
restored in this vacuum \cite{0007226,0008013,0009038}. It will be
interesting to analyse our
proposal to check these properties.
Another important problem is to generalize this analysis to the case
of superstring field theory (SSFT) \cite{9503099,9912120,9912121} where
analogous
conjectures
exist~\cite{9805019,9805170,9808141,9810188,9812031,9812135} and some
have already been tested using the level truncation
scheme \cite{0001084,0002211,0003220,0004015}. It is
again natural to conjecture that at the tachyon vacuum the kinetic term
does not involve any matter sector operator.

The paper is organised as follows. In section \ref{s2} we discuss the
properties that we expect of the open string field theory action expanded
around the tachyon vacuum, and construct examples of actions
satisfying
these properties. In section \ref{s3} we show how, starting from this
action, and assuming the existence of a space-time independent solution of
the equations of motion describing the original D-brane, we can calculate
the ratio of tensions of the lump solutions of different dimensions. In
the concluding section \ref{s4} we discuss various issues which are
important for further extension of our analysis. These include
a study of the
uniqueness of the proposed action, the possibility of using level truncation
scheme for calculating the overall normalization of the lump tension,
an analysis of gauge fixing,
generalization to superstrings, and a discussion on closed
string states.

\sectiono{String Field Theory  Around the Tachyon
Vacuum} \label{s2}

Here we begin by considering open string field theory formulated
on the background of a specific D-brane, focusing on  its algebraic
structure both before and after tachyon condensation to the
vacuum where the brane disappears.  After a discussion of the
properties expected from the kinetic term of the string field theory
around the tachyon vacuum we propose an ansatz consistent with
such expectations.

\subsection{Algebraic structure before and after condensation}
\label{s2.1}

We begin with the cubic string field theory action describing the
world-volume theory of a D-$(N-1)$-brane:
\be \label{e1}
S (\Phi) = \,-\, {1\over g_o^2}\,\,\bigg[\, {1\over 2} \langle \,\Phi
\,,\, Q_B\,
\Phi
\rangle + {1\over 3}\langle \,\Phi \,,\, \Phi *
\Phi \rangle \bigg] \,,
\ee
where $g_o$ is the open string coupling constant and $\Phi$ is the open
string field, conventionally taken to be Grassmann odd and of ghost
number one for the classical action. In addition, $Q_B$ is the
BRST operator, $\langle \cdot , \cdot \rangle$ is a bilinear inner
product based on BPZ conjugation and
$*$ denotes star-multiplication of string fields. The consistency
of this classical action is guaranteed by the following identities
involving the BRST operator
\ben \label{e1p}
&& Q_B^2 = 0, \nonumber \\
&& Q_B (A * B) = (Q_BA) * B + (-1)^{A} A * (Q_BB)\,, \\
&& \langle \, Q_B A , B \,\rangle = - (-)^A \langle A , Q_B B \rangle
\,,\nonumber
\een
and identities involving the inner product and the star operation
\ben
\label{e11p}
&& \langle A, B \rangle = (-)^{AB} \langle  B , A \rangle \nonumber \\
&& \langle \, A \,, B * C \, \rangle = \, \langle A* B \,,\, C \,
\rangle  \\
&& A * (B * C) = (A*B) * C \,.\nonumber
\een
In the sign factors, the exponents $A, B, \cdots$ denote the Grassmanality
of the state, and should be read as $(-)^A \equiv (-)^{\epsilon (A)}$
where $\epsilon (A) = 0 \, (\hbox{mod}~ 2)$ for $A$ Grassmann even,
and
$\epsilon (A) = 1 \, (\hbox{mod}~ 2)$ for $A$ Grassmann odd.  We also
have:
\ben \label{e1pp}
&& \epsilon (A * B ) = \epsilon (A) + \epsilon (B) \nonumber \\
&& \hbox{gh} (A* B ) = \hbox{gh} (A) + \hbox{gh} (B) \,,
\een
where gh denotes ghost number, and we take the ghost number of the
SL(2,R) vacuum to be zero.
Equations \refb{e1p} and \refb{e11p} guarantee that the above action is
invariant
under the gauge transformations:
\be \label{e5}
\delta \Phi = Q_B \Lambda + \Phi * \Lambda - \Lambda * \Phi\, ,
\ee
for any Grassmann-even ghost-number zero state $\Lambda$.

\bigskip
Let $\Phi_0$ be the string field
configuration describing the tachyon vacuum, a solution of the
classical field equations following from the  action in \refb{e1}:
\be
\label{feq}
Q_B \Phi_0 +  \Phi_0 * \Phi_0 = 0 \,.
\ee
If $\wt\Phi=\Phi -
\Phi_0$ denotes the shifted open string field, then
the cubic string field theory action expanded around the tachyon vacuum
has the form:
\be \label{e2}
S (\Phi_0 + \wt\Phi) = S(\Phi_0) \,-\, {1\over g_o^2}\,\,\bigg[\, {1\over
2} \langle
\,\wt\Phi
\,,\, Q\,
\wt\Phi
\rangle + {1\over 3}\langle \,\wt\Phi \,,\, \wt\Phi *
\wt\Phi \rangle \bigg] \,.
\ee
Here $S(\Phi_0)$ is a constant,
which according to the energetics
part of the tachyon conjectures equals the mass $M$
of the D-brane when the D-brane extends over a space-time of finite
volume.\footnote{As suggested in \cite{9912249}, we shall assume that the
time
interval has unit length so that the action can be identified with the
negative of the potential energy for static configurations.}
Indeed, the potential energy $V(\Phi_0) = - S(\Phi_0)$ associated to
this string field configuration should equal minus the mass of the
brane.
The kinetic operator $Q$ is given in terms
of $Q_B$ and $\Phi_0$ as:
\be \label{e3}
Q \wt\Phi = Q_B \wt\Phi + \Phi_0 * \wt\Phi + \wt\Phi* \Phi_0\, .
\ee
More generally, on arbitrary string fields one would define
\be \label{e3p}
Q A = Q_B A + \Phi_0 * A - (-1)^{A}  A * \Phi_0\, .
\ee
The consistency of the action \refb{e2}
is guaranteed from the
consistency of the one in \refb{e1}. Since neither the inner
product nor the star multiplication have changed,
the identities
in \refb{e11p} still hold. One can readily check that the identities
in \refb{e1p} hold when $Q_B$ is replaced by $Q$~\cite{WITTENBSFT}.
Just as \refb{e1} is invariant under the gauge trasformations
\refb{e5}, the action in \refb{e2} is invariant under
$\delta \wt\Phi = Q \Lambda + \wt\Phi * \Lambda - \Lambda * \wt\Phi$
for any Grassmann-even ghost-number zero state $\Lambda$.

Since the energy density of the brane represents
a positive cosmological constant,
it is natural to add the constant $-M=-S(\Phi_0)$ to
\refb{e1}. This will cancel the $S(\Phi_0)$ term in
\refb{e2}, and will make manifest
the
expected zero energy density in the final vacuum without D-brane. For the
analysis around this final vacuum it suffices therefore to
study the action
\be \label{e2p}
S_0 (\wt\Phi) \equiv \,-\, {1\over g_o^2}\,\,\bigg[\, {1\over 2} \langle
\,\wt\Phi
\,,\, Q\,
\wt\Phi
\rangle + {1\over 3}\langle \,\wt\Phi \,,\, \wt\Phi *
\wt\Phi \rangle \bigg] \,.
\ee

\subsection{An ansatz for the SFT after condensation} \label{s2.2}

If we had a closed form solution $\Phi_0$ available, the problem
of formulating SFT around the tachyon vacuum would be significantly
simplified, as we would only have to understand the properties of
the new kinetic operator $Q$ in \refb{e3p}. In particular we would
like to confirm that its cohomology vanishes in accordance with
the expectation that all conventional open string excitations disappear
in the tachyon vacuum.  Even the numerical approximations
for $\Phi_0$ may give some indication if this is the case
\cite{KS,0008033,ellwood}.

\medskip Previous experience with background deformations (small and
large) in SFT indicates that even if we knew $\Phi_0$ explicitly and
constructed $S_0(\wt\Phi)$ using eq.\refb{e2p}, this may not be the most
convenient form of the action.
Typically a nontrivial field
redefinition is
necessary to bring the shifted SFT action to the canonical form
representing the new background~\cite{9307088}.
In fact, in some cases, such as
in the formulation of open SFT for D-branes with various values of
magnetic fields, it is simple to formulate the various SFT's directly
\cite{9912274}, but the nontrivial classical solution relating
theories
with different magnetic fields are not known. This suggests that if a
simple form exists for the SFT action around the tachyon vacuum it might
be easier to guess it than to derive it.

In proposing a simple form of the tachyon action, we have in mind field
redefinitions of the action in \refb{e2p} that leave the cubic term
invariant but simplify the operator $Q$ in \refb{e3p} by transforming
it into a simpler operator $\QQ$.  To this end we consider homogeneous
field redefinitions of the type
\be
\label{jkl}
\wt\Phi = \, e^K \, \Psi\,,
\ee
where $K$ is a ghost number zero Grassmann
even operator. In addition, we require
\ben \label{efderp}
&& K (A * B) = (K A) * B +  A * (K B)\,, \nonumber  \\
&& \langle \, K A , B \,\rangle = - \langle A , K B \rangle \,.
\een
These properties guarantee that the form of the
cubic term is unchanged and
that after the field redefinition the action takes the form
\be \label{e2findp}
\SS (\Psi) \equiv \,-\, {1\over g_0^2}\,\,\bigg[\, {1\over 2} \langle
\,\Psi \,,\, \QQ\, \Psi
\rangle + {1\over 3}\langle \,\Psi \,,\, \Psi *
\Psi \rangle \bigg] \,,
\ee
where
\be\label{kjh}
\QQ =  \, e^{-K}  Q e^K\,.
\ee
Again, gauge invariance only requires:
\ben \label{eFINp}
&& \QQ^2 = 0, \nonumber \\
&& \QQ (A * B) = (\QQ A) * B + (-1)^{A} A * (\QQ B)\, , \\
&& \langle \, \QQ A , B \,\rangle = - (-)^A \langle A , \QQ B \rangle
\,. \nonumber
\een
These identities hold
by virtue of \refb{efderp} and \refb{kjh}.
We will proceed here postulating a $\QQ$ that satisfies these
identities as well as other conditions, since lacking knowledge
of $\Phi_0$ the above field redefinitions cannot be attempted.

\bigskip
The choice of $\QQ$ will be required to satisfy the following properties:
\begin{itemize}

\item
The operator $\QQ$ must be of ghost number one
and must satisfy the conditions
\refb{eFINp}
that guarantee gauge invariance of the string action.

\item  The operator $\QQ$ must have vanishing cohomology.

\item The operator $\QQ$ must be universal, namely, it must be possible to
write without reference to the brane boundary conformal field theory.

\end{itemize}

The simplest possibility would be to set $\QQ=0$. This would result in the
purely cubic version of open string field theory \cite{CUBIC}. Indeed, it
has
long been tempting to identify the tachyon vacuum with a theory where
the kinetic operator vanishes, especially because lacking the kinetic term
the string field gauge symmetries are not spontaneously broken.
Nevertheless, there are well-known complications with this identification.
At least in the construction of \cite{CUBIC} the string field shift
$\overline
\Phi$ relating the cubic to the purely cubic SFT's involves the subtle
identity operator ${\cal I}$ and appears to satisfy $Q \overline \Phi=0$
as well as $\overline \Phi*\overline \Phi=0$. The tachyon condensate
definitely does not satisfy these two identities. We therefore search for
nonzero $\QQ$.

We can satisfy the three requirements by letting
$\QQ$ be constructed purely from ghost operators. In particular we claim
that the ghost number one operators
\be
\label{cn}
\CC_n \equiv  c_n +  (-)^n \, c_{-n}  \,, \quad n=0,1,2,\cdots
\ee
satisfy the properties
\ben \label{ecp}
&& \CC_n \CC_n  = 0, \nonumber \\
&&\CC_n (A * B) = (\CC_n A) * B + (-1)^{A} A * (\CC_n B)\,, \\
&& \langle \, \CC_n A , B \,\rangle = - (-)^A \langle A , \CC_n B \rangle
\,.\nonumber
\een
The first
property is manifest.
The last property
follows
because under BPZ conjugation $c_n \to (-)^{n+1} c_{-n}$.
The second property
follows from the
conservation law
\be
\langle V_3| (\CC_n^{(1)} + \CC_n^{(2)} + \CC_n^{(3)} ) = 0\,,
\ee
on the three string vertex \cite{0006240}. These conservation
laws arise from consideration of integrals of the form
$\int dz  c(z) \varphi(z)$ where $\varphi(z) (dz)^2$ is a globally
defined quadratic differential.  In fact, the conservation law for
$\CC_0$ arises from a familiar Jenkins-Strebel quadratic differential
with a second order pole at each of the three puntures. This quadratic
differential, in fact, defines the geometry of the vertex. Its
horizontal trajectories show the open strings and their interaction.

Each of the operators $\CC_n$ has
vanishing cohomology since for each
$n$ the operator $B_n ={1\over 2} (b_n +  (-)^n \, b_{-n}) $ satisfies
$\{ \CC_n , B_n \} = 1$.  It then follows that whenever $\CC_n \psi=0$, we
have $\psi = \{  \CC_n , B_n \} \psi = \CC_n ( B_n \psi)$, showing that
$\psi$ is  $\CC_n$ trivial.
Finally, since they are built from ghost oscillators, all $\CC_n$'s
are manifestly universal.

It is clear from the structure of the conditions \refb{eFINp}
that they are satisfied for the general choice:
\be
\label{choice}
\QQ = \sum_{n=0}^\infty a_n \, \CC_n \,,
\ee
where the $a_n$'s are constant coefficients. We will discuss
in section \ref{s4.1}
how different choices for these constants result
in kinetic operators that are sometimes
related by field redefinitions.

A subset of the above kinetic operators were considered by Horowitz
{\it et.al.}\cite{HORO}, as they discussed how the purely cubic SFT
could have
solutions leading to forms different from the conventional cubic SFT. In
particular they noticed that operators built just with ghost oscillators
could appear as the kinetic term of an action obtained by shifting by a
(formal) solution of the cubic theory.

There may be other choices of $\QQ$ satisfying all the requirements stated
above. Fortunately, our analysis of section 3 will not
require the knowledge of
the detailed form of $\QQ$,
as long as it does not involve any matter
operators. To this end,
it will be useful to note that since $\QQ$ does
not involve matter operators, we can fix the gauge by choosing a gauge
fixing condition that also does not involve any matter operator. In such a
gauge, the propagator will factor into a non-trivial operator in the ghost
sector, and the identity operator in the matter sector.
Some details of gauge fixing are discussed
in section \refb{s4.3}.

\sectiono{The Lump Solutions and Their Tensions} \label{s3}

In this section we shall analyze classical solutions of the string field
action \refb{e2findp}
introduced in the previous section.
The analysis will proceed in three steps: 
1) analysis of the tachyon
effective
action, 2) construction of the lump solution, 
and 3) computation of the ratio
of the tensions of the lump solutions.

\subsection{Analysis of the tachyon effective action} \label{s3.1}

We shall consider a situation where we start with an unstable D-$(N-1)$
brane of the bosonic string theory, and consider the string field theory
action expanded around the tachyon vacuum of this theory. According to
our postulates of the previous section this action will have the form
given in eq.\refb{e2findp}.
Given this action
and a suitable gauge fixing which does
not involve any matter operator, the propagator involves purely ghost
sector operators. Since both
the vertex and the propagator now factorize into matter sector
contribution and ghost
sector contribution, a general $N$-point function will also factorize into
a matter
sector contribution and a ghost sector contribution.

We shall now concentrate on the $n$-tachyon
Green's function. If we
denote by $J(p)$ the current coupling to a tachyon of momentum
$(-p)$,
then the
generating functional of the tachyon Green's functions has the form:
\be \label{e7}
W[J] = \sum_{n=2}^\infty {1 \over n!} \int d^Np_1 \ldots d^Np_n
g^{(n)}(p_1, \ldots p_n) J(p_1)\ldots J(p_n) \delta(\sum_{i=1}^n p_i) \, .
\ee
Here $g^{(n)}(p_1, \ldots p_n)\delta(\sum_{i=1}^n p_i)$
is the  $n$-tachyon off-shell amplitude obtained by
adding all tree level connected Feynman diagrams. Note that since we
consider
a propagator that involves just simple ghost factor,
its world sheet interpretation
is that of gluing with a zero length strip and inserting a ghost operator.
Thus the propagator collapses to an overlap with a
ghost insertion.\footnote{If the ghost kinetic operator does not
have a simple inverse the fact remains that in the matter sector
the propagator will act as an overlap.}
Recall that the
cubic vertex corresponds to the symmetric gluing of three semi-infinite
strips
across the open edges, half edges at a time.  It follows that for $n$
external
tachyons any connected Feynman graph constructed with this vertex and the
overlap propagator will simply be built by the symmetric gluing of
$n$-semiinfinite strips across their open edges, half edges at a time.
Mapping this world sheet into a unit disk, each strip turns into
a wedge of angle $2\pi/n$.  Except for the ghost insertions this can be
viewed as the iterated star product of $n$
tachyon vertex operators. In fact, such
iterated products are precisely the ones used in the WZW-like superstring
field theory \cite{9503099} as discussed in detail in \cite{0002211}.

It follows from the above remarks that the $n$-tachyon off
shell amplitude can be written as
the following correlation function on a unit disk:
\be \label{e19}
g^{(n)}(p_1,\ldots p_n) \delta(\sum_{i=1}^n p_i) = C_n \langle
\prod_{k=1}^n (f^{(n)}_k \circ e^{i
p_k\cdot X(0)})\rangle\, ,
\ee
where $\langle\cdots\rangle$ denotes matter sector correlation
function, $C_n$ is a constant representing contribution from the ghost
sector correlator,
$f^{(n)}_k \circ e^{i p_k \cdot X(0)}$ denotes the conformal transform
of $e^{i p_k\cdot X(0)}$ by the function $f^{(n)}_k(z)$:
\be \label{e21}
f^{(n)}_k \circ e^{i p_k \cdot X(0)} = |f^{(n)\prime}_k(0)|^{p_k^2}e^{i
p_k
\cdot X(f^{(n)}_k(0))}\, ,
\ee
and \cite{0002211}:
\be \label{e20}
f^{(n)}_k(z) = e^{2\pi i (k-1)/n} \Big({1 + i z \over 1 -
iz}\Big)^{2/n}\, .
\ee
Using the relations:
\be \label{e22}
f^{(n)}_k(0) = e^{2\pi i (k-1)/n}, \qquad f^{(n)\prime}_k(0) = e^{2\pi i
(k-1)/n} {4\over n} i\, ,
\ee
and
\be \label{e23}
\langle \prod_{k=1}^n  e^{i
p_k\cdot X(f^{(n)}_k(0))}\rangle = \delta(\sum_{k=1}^n p_k)
\prod_{k \ne l}
|f^{(n)}_k(0) -
f^{(n)}_l(0)|^{p_k\cdot  p_l} \, ,
\ee
we get
\be \label{e24}
g^{(n)}(p_1,\ldots p_n) =
C_n \exp\Big[\,\Bigl(\ln{4\over n}\,\Bigr) %bz ( ) for readability
\,{\sum_{k=1}^n p_k^2
+\sum_{k, l =1\atop k\ne l}^n
p_k\cdot p_l  \ln\Bigl( 2\sin\,
({\pi\over n}|k-l|\,)\Bigr)}
\Big]\, , \quad \hbox{for} \quad n\ge 3\, .
\ee
While $g^{(n)}$ given above is cyclically symmetric in
$p_1,\ldots p_n$,
only the fully symmetric part of $g^{(n)}$
contributes to $W$. The tachyon propagator $g^{(2)}$
is a momentum independent constant.

\medskip
We now perform a series of manipulations familiar in path integral
quantization of field theories.
The tachyon
effective action can be obtained from the generating functional $W$ by
taking the
Legendre transform. As usual, we introduce the classical
field expectation value as
\be \label{en1}
\phi(q) = \phi_c[q, J]\, ,
\ee
where
\be \label{e10}
\phi_c[-p_1, J] \equiv {\delta W \over \delta J(p_1)} =
\sum_{n=2}^\infty {1
\over (n-1)!} \int d^Np_2 \ldots d^Np_n
g^{(n)}(p_1, \ldots p_n) \delta(\sum_{i=1}^n p_i) J(p_2)\ldots
J(p_n)\, .
\ee
Note that $\phi_c[q,J]$ is a function of
the momentum variable
$q$ and a functional of $J(p)$.
In deriving the above equation we have used the cyclic symmetry of
$g^{(n)}$.
The effective action, a functional of classical fields
$\phi(q)$,  is
given
by,
\ben \label{e8a} \Gamma[\phi]
&=& \int d^Np J(p) \phi(-p) - W[J]\,, \een
where $J$ above is the source
that gives rise  to $\phi(q)$ through \refb{en1}, \refb{e10}.
Since $g^{(2)}$ is momentum independent, the
coefficient of the quadratic term in $\Gamma[\phi]$ will be momentum
independent and the higher order terms will have exponential dependence on
the momenta. This resembles the action of the  $p$-adic string theory
expanded
around the tachyon vacuum\cite{BREKKE,0003278} after a momentum
dependent
rescaling of the tachyon field. As in $p$-adic string theory we shall
see that gaussians play an important role in constructing lump
solutions.

It follows from equation \refb{e8a} that
\be \label{ejp}
J(-p) = {\delta \Gamma[\phi]\over \delta\phi(p)}\, .
\ee
Thus,
a solution to the
equations of motion arising from
setting the variation of the effective action to zero
amounts to setting the current $J$ to zero. Therefore, classical
solutions are of the form:
\be \label{e11}
\phi(q) = \phi_c[q, J(p)=0] .
\ee

Since $\phi_c[q, J]$ has a Taylor series expension in $J$ starting with
linear
power, this would seem to always give the trivial solution $\phi (q) =0$.
However, $W$ and $\phi_c$ have branch points in the complex
$[J(p)]$
space, and by going around these branch points, one may get $\phi_c[q,
J(p)=0]
\not=0$.
Every such path in the $J$ space will give rise to a
valid solution of the equations of motion.\footnote{The situation
can be
illustrated by considering the zero momentum sector of the $\phi^3$ field
theory. In this case, at the tree level, $\Gamma[\phi] = -{1\over 2} m^2
\phi^2 + {1\over 3} \lambda \phi^3$. This gives $J =
-m^2\phi
+ \lambda\phi^2$, and $\phi = {1\over 2\lambda} (m^2
\pm \sqrt{m^4
+ 4
\lambda
J}) \equiv \phi_c[J]$. If we choose the $-$ sign in front of the square
root,
$\phi_c[J]$
vanishes at $J=0$ and has a Taylor series expansion in $J$ starting at
linear order. But going around the branch point at $J=-m^4/(4\lambda)$ in
the complex $J$ plane, we can move to the other branch with $+$ sign in
front of the square root. Now if we return to the origin $J=0$, we get
$\phi_c[0]=m^2/\lambda$. This is indeed the location of an extremum of
$\Gamma[\phi_c]$ and hence represents a solution of the equations of
motion.}

\subsection{Setup for lump solutions} \label{s3.2a}

We shall begin by examining translationally invariant solutions of the
equations of motion.
(The original D-$(N-1)$-brane
configuration
will be represented by such a solution.) In momentum space a
translationally
invariant solution
$\phi(p)$ is proportional to $\delta(p)$
where $p$
is an $N$-dimensional momentum vector.
Consider eq.~\refb{e10} with a delta function
source $J(p) = u \delta(p)$.
We find
\be \label{e13}
\phi_c [q, \uu \delta(p)]\equiv
F(\uu) \delta(q)\, ,
\ee
with
\be \label{e15}
F(\uu) = \sum_{n=2}^\infty {1 \over
(n-1)!} \uu^{n-1}
g^{(n)}(0, \ldots 0) = \sum_{n=2}^\infty {1 \over
(n-1)!} C_n \uu^{n-1}\, ,
\ee
where use was made of \refb{e24} for zero momentum.
Now suppose that the function
$F(\uu)$
has a branch point $\uu_b$ in the complex $\uu$ plane such that as
$\uu$
returns to 0 after going around this branch point, $F(\uu)$ returns to
$\phi_0$. Thus in this branch
\be
\lim_{\uu\to 0'} F(u)=\phi_0\,,
\ee
where $0'$ is used to denote the origin on this branch. Hence
$\lim_{\uu\to 0'}\phi_c (q,\uu\delta(p))=\phi_0\delta(q)$.
Eq.\refb{e11} then
implies the
existence of a translationally invariant
solution of the equations of motion of
the form
\be \label{e12}
\phi(p) = \phi_0 \delta(p)\, .
\ee
Conversely, existence of a solution of the form given in \refb{e12}
implies that the function $F(\uu)$ has a branch point $u_b$
in the
complex
$\uu$ plane such that as $\uu$
returns to 0 after going around the branch point, $F(\uu)$ returns to
$\phi_0$. We shall choose $\phi_0$ to be the value that describes the
original D-brane solution before tachyon condensation;
such a solution exists by assumption.
In what follows, we shall assume that
the branch point $\uu_b$ is the closest singularity to the origin in the
complex $\uu$ plane, so that the growth of the coefficients in the
Taylor series expansion of $F(\uu)$ is controlled by $\uu_b$.
In this case, the radius of convergence of the Taylor
series expansion of $F(u)$ given in eq.\refb{e15} is $|u_b|$.

Now suppose we find a pair of functions $\psi(p)$ and $\chi(p)$
such
that,
\be \label{e14}
H(\uu, q) \equiv \phi_c\,[q,
\uu\psi(p)] - F(\uu) \chi(q)
\ee
has the property that the rate of growth of the coefficients in the Taylor
series expansion of $H(\uu, q)$ in $\uu$ is slower than that of
$F(\uu)\chi(q)$. In this case we can conclude that
\begin{enumerate}
\item $H(\uu, q)$ does not have a singularity for $|\uu|<|\uu_b|$.
\item $H(\uu, q)$ either has no singularity at $\uu=\uu_b$ or its
singularities at $\uu=\uu_b$ are non-leading compared to that of
$F(\uu)\chi(q)$.
\end{enumerate}
We can now consider tracing the path around $\uu_b$ in the complex
$\uu$-plane so that $F(\uu)$ approaches $\phi_0$ as $\uu$ returns to
the origin. If $H(\uu, q)$ has no singularity at $\uu=\uu_b$, then
it will return to zero. Otherwise it will return to some function
$h(q)$
\be
\lim_{\uu\to 0'} H(u, q)=h(q)\, .
\ee
As a result, on this
branch \refb{e14} gives:
\be \label{e14a}
\lim_{\uu\to 0'} \phi_c\,[q,\uu \psi(p)] = \phi_0 \chi(q) + h(q)\, .
\ee
Eq.\refb{e11} then implies that
\be
\label{iasol}
\phi(q) = \phi_0\chi(q)+h(q)\,,
\ee
is a solution of the
equations of motion. Thus given two functions $\psi(q)$ and $\chi(q)$
satisfying the property
stated below eq.\refb{e14}, and a translationally invariant solution
$\phi_0\delta(q)$, we can
find a space-time dependent solution $\phi_0\chi(q)+h(q)$.
Note that we have defined $\chi(q)$ to be such that $F(u)\chi(q)$ is
the leading contribution to $\phi_c (q, u\psi(p))$; so by definition, the
left-over piece $H$ is sub-leading i.e. its power series expansion
should converge faster than that of $F(u)\chi(q)$.
However, the existence
of such a $\chi(q)$ is not guaranteed {\it a priori} since the
$u$-dependence of $\phi_c(q, u\psi(p))$ could be very different from that
of $\phi_c(q, u\delta(p))$.

\medskip
We shall now
make the convergence condition precise. Let us define
\be \label{e17}
R_n(-p_1) \equiv {
\int d^Np_2 \ldots d^Np_n
g^{(n)}(p_1, \ldots p_n) \delta(\sum_{i=1}^n p_i) \psi(p_2)\ldots
\psi(p_n) \over  g^{(n)}(0, \ldots 0) \chi(-p_1)}\, .
\ee
Making use of this definition, and of
eqs.~\refb{e10},
\refb{e15} and \refb{e17}, equation \refb{e14} gives
\be \label{e16}
H(u, -p) =  \chi(-p) \sum_{n=2}^\infty {u^{n-1}
\over (n-1)!} C_n  [\,R_n(-p) -1 \,]\,.
\ee
We need to ensure that the Taylor series expansion of $H(\uu,q)$
converges more
rapidly than the Taylor series expansion of $F(\uu)\chi(q)$ given in
eq.\refb{e15}. This, in
turn
requires that
\be \label{e17a}
\lim_{n\to\infty} R_n(-p) = 1\, .
\ee
Note that the dependence on the constants $C_n$ arising from
ghost correlators in \refb{e24} drops out for $R_n$ defined in
eq.\refb{e17}. Equation
\refb{e17a} will be the key equation in the analysis that follows.

\subsection{Equations for lump solutions} \label{s3.2}

In order to proceed further, we need to have an ansatz for $\psi(p)$ and
$\chi(p)$.
We use the following ansatz:
\be \label{e18}
\psi(p) = K \exp({-\alpha p_\perp^2 / 2}) \delta(p_\parallel)\, ,
\qquad \chi(p) =
\gamma K \exp({-\beta p_\perp^2 / 2}) \delta(p_\parallel)\, ,
\ee
where $K$, $\alpha$, $\beta$ and $\gamma$ are constants to be determined,
and
$(p_\parallel,
p_\perp)$ denotes a decomposition of the vector $p$ into two orthogonal
subspaces of dimensions $N_\perp$ and $N_\parallel$ respectively
($N_\perp+N_\parallel=N)$. In position space labelled by
$(x_\parallel, x_\perp)$ this corresponds to a configuration
which is independent of $x_\parallel$ and has the form of a
gaussian along $x_\perp$. Thus a natural interpretation of this
solution is that of a solitonic ($N_\parallel -1)$ brane along
$x_\parallel$.
Substituting \refb{e18}
into eq.\refb{e17}, 
we get 
\ben \label{eneweq}
&& \gamma\, K\, R_n(-p_1)
g^{(n)}(0, \ldots 0)  \exp({-\beta p_{1\perp}^2 /
2})\delta(p_{1\parallel}) \nonumber \\
&=& K^{n-1}\int d^Np_2 \ldots d^Np_n
g^{(n)}(p_1, \ldots p_n) \delta(\sum_{i=1}^n p_i) 
\exp\Big(-{\alpha\over 2}\sum_{i=2}^n p_{i\perp}^2\Big)\prod_{i=2}^n
\delta(p_{i\parallel})\, .
\een
Using \refb{e24} in eq.\refb{eneweq}, performing
the integrals over $p_{i\parallel}$, with $i=2, \cdots n$, 
and using the freedom of setting $p_{1\parallel}=0$ 
in $g^{(n)}(p_1,\ldots p_n)$ and $R_n(-p_1)$ due to the
accompanying factors of $\delta(p_{1\parallel})$
we get  
\be \label{e25}
\gamma R_n(-p_{1\perp}) 
 \exp\Bigl({-{ (\alpha+\beta)\over 2}
p_{1\perp}^2}\Bigr) = K^{n-2}\hskip-3pt \int
\hskip-3pt d^{N_\perp} p_{2\perp} \ldots
d^{N_\perp} p_{n\perp}
\delta(\sum_{i=1}^n p_{i\perp})
\exp\Big({\sum_{k,l=1}^n a_{k-l} p_{k\perp} \cdot p_{l\perp}}\Big)\, ,
\ee
where
\ben \label{e26}
a_{0} &=& \ln{4\over n} - {\alpha\over 2} \,, 
\nonumber \\
a_k &=& \ln\Bigl| 2\sin({\pi k \over n})\Bigr| \quad \hbox{for} \quad
k\ne 0
\, .
\een
We now multiply both sides of the equation by $e^{ip_{1\perp}\cdot
x_\perp}$ for some $N_\perp$ dimensional vector $x_\perp$ and integrate
over $p_{1\perp}$. Demanding that in the $n\to\infty$ limit $R_n$
should approach $1$, we get
\ben \label{e27}
 &&\hskip-20pt \gamma \Big({2\pi\over \alpha+\beta}\Big)^{N_\perp / 2}
\exp({-{x_\perp^2\over 2(\alpha+\beta)}}) \nonumber \\
 = \lim_{n\to\infty} &&\hskip-20pt\Big[K^{n-2} \int \prod_{k=1}^n
d^{N_\perp}p_{k\perp}
\delta(\sum_{i=1}^n p_{i\perp}) \exp\Big({ip_{1\perp}\cdot
x_\perp} + {\sum_{k,l=1}^n a_{k-l} p_{k\perp} \cdot p_{l\perp}}\Big)
\Big] \, .
%\nonumber \\
\een
In order to do the integral over $p_{i\perp}$ on the right hand side, we
take help of the discrete Fourier transform. We introduce new variables
$\phi_{l\perp}$ as follows:
\be \label{e28}
\phi_{l\perp} = {1\over \sqrt n} \sum_{k=1}^n p_{k\perp} e^{-2\pi i (k-1)
l / n}\, , \qquad 1\le l \le n\, .
\ee
Then
\be \label{e29}
p_{k\perp} = {1\over \sqrt n} \sum_{l=1}^{n} \phi_{l\perp} e^{2\pi i
(k-1) l / n}\, .
\ee
The reality condition on $\phi_{l\perp}$ is of the form:
\be \label{e30}
\phi_{l\perp}^* = \phi_{(n-l)\perp}\quad \hbox{for} \quad 1\le l\le
(n-1), \qquad \phi_{n\perp}^*=\phi_{n\perp}\, .
\ee
Let us first take $n$ to be odd. In that case we can
introduce a new set of variables $\xi_{s\perp}$, $\eta_{s\perp}$ through
the relations:
\be \label{e31}
\phi_{s\perp} = {1\over \sqrt 2} (\xi_{s\perp} + i \eta_{s\perp})\,
\qquad
\hbox{for} \quad 1\le s \le {n-1\over 2}\, .
\ee
We take $\phi_{n\perp}$, $\xi_{s\perp}$ and $\eta_{s\perp}$ as
independent variables, and express the right hand side of eq.\refb{e27} in
terms of these variables. This gives
\ben \label{e32}
&&\gamma \Big({2\pi\over \alpha+\beta}\Big)^{N_\perp / 2}
\exp({-{x_\perp^2\over 2(\alpha+\beta)}})
\nonumber
\\
&=& \lim_{n\to\infty} \bigg[
K^{n-2} \int d^{N_\perp}
\phi_{n\perp}
\prod_{s=1}^{(n-1)/2} (d^{N_\perp}\xi_{s\perp}d^{N_\perp}\eta_{s\perp})
\delta(\sqrt n \phi_{n\perp}) \nonumber \\
&& \exp\Big({{i\over \sqrt n} (\phi_{n\perp} +
\sqrt
2\sum_{s=1}^{(n-1)/2} \xi_{s\perp}) \cdot
x_\perp} - b_n \phi_{n\perp}^2 - \sum_{s=1}^{(n-1)/2} b_s
(\xi_{s\perp}^2 + \eta_{s\perp}^2)
\Big)\bigg]\, ,
\een
where
\be \label{e33}
b_s = -\sum_{k=0}^{n-1}
a_k \cos\Bigl( {2\pi k s\over n}\Bigr) =
\ln{n\over 4} + {\alpha\over 2} - \sum_{k=1}^{n-1} \ln
\Bigl(2\sin({\pi k \over n})\Bigr)
\cos\Bigl( {2\pi k s\over n}\Bigr)\, . \ee
We can now perform the integrals over $\phi_{n\perp}$, $\xi_{s\perp}$ and
$\eta_{s\perp}$ explicitly. This gives
\be \label{e34pre}
\gamma\Big({2\pi\over \alpha+\beta}\Big)^{N_\perp / 2}
\exp({-{x_\perp^2\over 2(\alpha+\beta)}})
= \lim_{n\to\infty} \Big[K^{n-2} \Big\{{1\over \sqrt n}
\Big(\prod_{s=1}^{(n-1)/2}
{\pi\over b_s}\Big)\Big\}^{N_\perp} \exp\Big(-{1\over 2n} x_\perp^2
\sum_{s=1}^{(n-1)/2}{1\over
b_s}\Big)\Big]\, .
\ee
Using the symmetry $b_s=b_{n-s}$, we can rewrite this equation as
\be \label{e34}
\gamma\Big({2\pi\over \alpha+\beta}\Big)^{N_\perp / 2}
\exp({-{x_\perp^2\over 2(\alpha+\beta)}})
= \lim_{n\to\infty} \Big[K^{n-2} \Big\{{1\over n}
\Big(\prod_{s=1}^{n-1}
{\pi\over b_s}\Big)\Big\}^{N_\perp/2} \exp\Big(-{1\over 4n} x_\perp^2
\sum_{s=1}^{n-1}{1\over
b_s}\Big)\Big]\, .
\ee
For $n$ even, we take the independent real variables to be
$\phi_{n\perp}$, $\phi_{{n\over 2}\perp}$, $\xi_{s\perp}$ and
$\eta_{s\perp}$ for $1\le s\le (n-2)/2$.
Repeating the analysis, we arrive at exactly the same
equation \refb{e34}. Thus eq.\refb{e34} is valid irrespective of whether
we take the limit $n\to\infty$ keeping $n$ even or odd. This is the main
equation that must be analyzed.

\subsection{Solving the lump equations} \label{s3.3}

We shall now try to examine if we can find solutions
of \refb{e34}
where
$\alpha$, $\beta$, $\gamma$ and $K$ are finite.
To this end let us define
the functions:
\be \label{es2a}
f(\alpha, n) = \sum_{s=1}^{n-1} \ln{\pi\over b_s} - \ln n\, , \qquad
k(\alpha, n) = {1\over n} \sum_{s=1}^{n-1} {1\over b_s}\, .
\ee
Comparing the two sides of
equation \refb{e34} we find that
\be \label{e35}
\beta = \lim_{n\to\infty} {2 \over k(\alpha, n)} - \alpha \equiv
C(\alpha)\, ,
\ee
\be \label{e36}
\lim_{n\to\infty} \Big[ (n-2) \ln K + {1\over 2} N_\perp
f(\alpha, n) - \ln \gamma - {1\over 2} N_\perp
\ln{2\pi\over \alpha + \beta} \Big] = 0\, .
\ee
Equation \refb{e35} arises from comparing the coefficients of the
gaussians in $x_\perp$ in the two sides of equation \refb{e34}. These
coefficients must coincide.  Once the gaussians coincide, the prefactors
must coincide as well.
Eqn. \refb{e36} arises by taking the logarithms of the
prefactors in \refb{e34}.
\begin{figure}[!ht]
\leavevmode
\begin{center}
\epsfbox{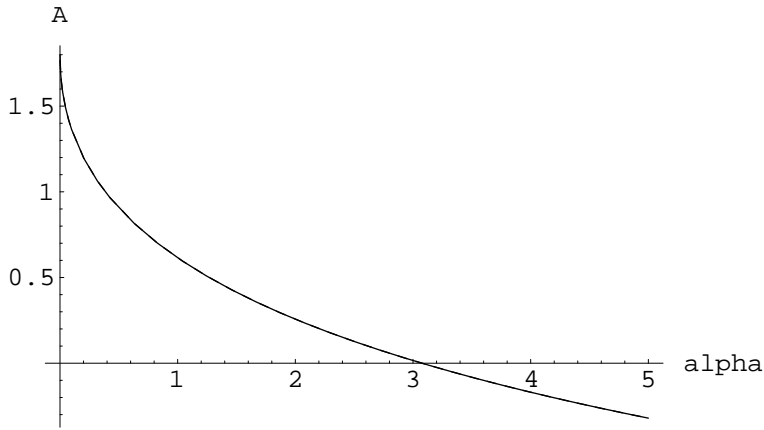}
\end{center}
\caption[]{\small The function $A(\alpha)$ for $n=50$ (dashed)
and $n=500$ (continuous). Actually the error is so
small that the lines are essentially on top of each other.} \label{f1}
\end{figure}

\begin{figure}[!ht]
\leavevmode
\begin{center}
\epsfbox{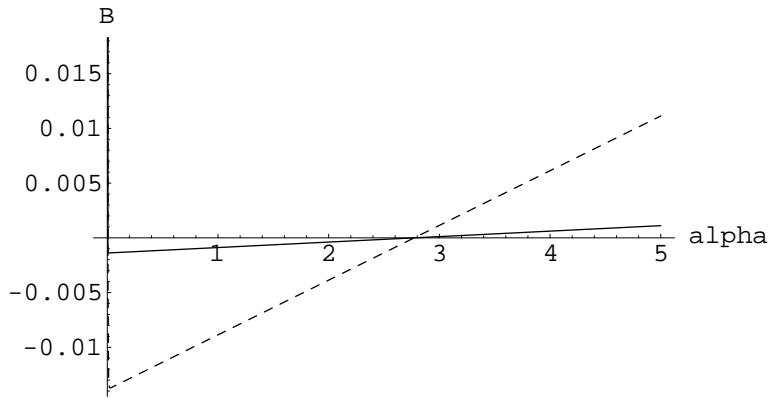}
\end{center}
\caption[]{\small The function ${\cal B} (n,\alpha)$
(equation \refb{cB}) plotted
for $n=50$ (dashed)
and $n=500$ (continuous).} \label{f2}
\end{figure}

\begin{figure}[!ht]
\leavevmode
\begin{center}
\epsfbox{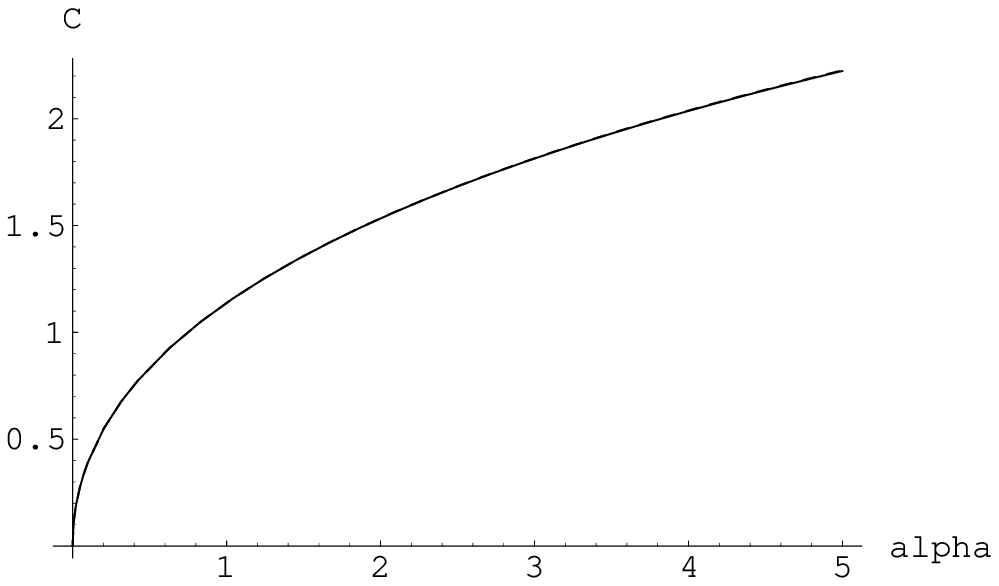}
\end{center}
\caption[]{\small The function $C(\alpha)$ (eqn.~\refb{e35}) for $n=50$
(dashed) and $n=500$ (continuous).
Actually the error is so
small that the lines are essentially on top of each other.} \label{f3}
\end{figure}

Numerical analysis shows that the function $k(\alpha, n)$
defined through
eq.\refb{es2a}
remains finite in the $n\to\infty$ limit.
Hence $C(\alpha)$ is well defined.
Furthermore, $f(\alpha, n)$
defined in eq.\refb{es2a} has the form
\be \label{es2}
f(\alpha, n) = A(\alpha) n + B(\alpha) + s_n(\alpha)\, ,
\ee
where $s_n(\alpha)$ vanishes in the $n\to\infty$ limit. In fact the
function $B(\alpha)$ appears to vanish identically,
but we shall for the
time being, not explicitly set it to zero.
We show in figure~\ref{f1}
a plot of $A(\alpha)$ calculated
numerically as $f(\alpha, n+1) - f(\alpha, n)$
for $n=50$ (in dashed
lines) and for $n=500$ (in continuous lines). Actually the error is so
small that the lines are essentially on top of each other.
In figure~\ref{f2} we
plot the quantity
\be
\label{cB}
{\cal B} (n, \alpha) \equiv 2 f(\alpha, n) - f(\alpha, 2n) =  B(\alpha) +
2 s_n(\alpha) - s_{2n}(\alpha) \,.
\ee
With $s_n \to 0$ as $n\to \infty$, for sufficiently large $n$ we have
that ${\cal B} (n, \alpha)$
approaches $B(\alpha)$. Thus a nonzero  $B(\alpha)$ would show as a constant
function in the plot of ${\cal B} (n, \alpha)$. Nevertheless, the numerical
evidence is that ${\cal B} (n, \alpha)$ goes to zero accurately as $n$ is
made large. Indeed, as can be seen in figure~\ref{f2}
even for $n=50$ (dashed
line) ${\cal B} (n, \alpha)$ is quite small over the considered range. When
$n=500$  (continuous line) ${\cal B} (n, \alpha)$ becomes much smaller.
We have explored this vanishing taking $n\sim 5000$ and have concluded that
with very high accuracy (about one part in ten thousand)
\be
\label{bzero}
B(\alpha ) \simeq 0\,,
\ee
for the range of $\alpha$ shown in the figure.
Having concluded  that $B(\alpha)=0$, the graph of ${\cal B} (n,
\alpha)$ gives us information about the $s_n(\alpha) $ terms. We have found
that the leading $s_n(\alpha)$ term is of order $\OO (1/n)$.
With $s_n(\alpha)
\sim 1/n$, and for large $n$ we have that
${\cal B} (n, \alpha) \simeq 3 s_n(\alpha) / 2 $
(see \refb{cB}).  Thus
figure~\ref{f2} is giving us  a plot of the leading $s_n(\alpha)$ term.
Note,
that there is a value $\bar\alpha$ for which
${\cal B} (n, \bar\alpha) =0$.  We determined numerically this value to be
$\bar\alpha
\simeq 2.772588$ and believe that it corresponds exactly to
$\bar\alpha = 4 \ln (2)$. Moreover the dependence of $s_n (\alpha)$
on $\alpha$ seems linear to high accuracy for a large range
around $\bar\alpha$ (see figure).
We have determined that
\be
\label{e342}
s_n (\alpha)   \simeq  {1\over 2} \,\, {0.3333\over n} \,(\alpha - \bar
\alpha) +
\OO (1/n^2)\,
\,,
\qquad  \bar \alpha = 4 \ln (2) \,.
\ee
The coefficient $0.3333$ may really take the exact value of
$1/3$ but we do
not know for certain. Our investigations also suggest that for
$\alpha=
\bar \alpha$ the $1/n^2$ corrections vanish while the $1/n^3$ corrections
are nonzero. It is hoped that these observations may help develop
an analytical approach to this large $n$ problem.
Finally in
figure~\ref{f3} we show the function $C(\alpha)$. This concludes our
discussion of the properties of \refb{es2}.

\bigskip
Substituting \refb{es2} into \refb{e36} and comparing terms linear in $n$
as well as the constant terms in both sides of the
equation we get:\footnote{Note that the existence of $K$ and $\gamma$
satisfying eq.\refb{e36} depends crucially on the asymptotic form
\refb{es2} of $f(\alpha, n)$. This, in turn, is a non-trivial
consequence of the specific form of the
$n$-tachyon vertex given in eq.(\ref{e24}). For example, if $g^{(n)}$ had
been momentum independent then we would get $b_s=\alpha/2$ and
$f(\alpha, n)=(n-1) \ln(2\pi/\alpha) - \ln n$. This does not
have the form \refb{es2}.}
\ben \label{es3}
\ln K &=& -{1\over 2} N_\perp A(\alpha)\,,
\nonumber \\
\ln\gamma &=& {1\over 2} N_\perp
\bigg(- \ln{2\pi\over \alpha+\beta} +
B(\alpha) + 2 A(\alpha)\bigg)\,.
\een
For a given value of $\alpha$, the above, together with $\beta = C(\alpha)$
(eqn.~\refb{e35}) determine the functions $\psi(p)$ and $\chi(p)$ 
through eq.~\refb{e18}. These functions, in turn,
generate a solution of the equations of motion of the form
$\phi_0\chi(p) + h(p)$ (eqn~\refb{iasol}).  

As already pointed out, the natural
interpretation of this solution is that of a $(N_\parallel -1)$-brane; we shall
refer to this as
a solitonic $(N_\parallel -1)$-brane. We would like to identify this with
D-($N_\parallel -1)$-brane.
However, since there seems to be a one parameter family of solutions
labelled by the parameter $\alpha$, we have the embarassment of
riches.
We shall now argue that the full solution $\phi_0\chi(p)+h(p)$ is
independent of the choice of $\alpha$.
To this end, note that using different values of $\alpha$ corresponds to
moving along different paths in the complex
$J(p)$ plane and returning to the origin.
Indeed, we are considering the subspace of the full
$J(p)$ space characterized by the parameters $\alpha$ and $u$ as follows:
\be \label{eneweqn}
J(p) = u e^{- N_\perp A(\alpha) / 2} e^{- \alpha
p_\perp^2 / 2}
\delta(p_\parallel) \, .
\ee
In the complex $(u, \alpha)$ plane we go around $u=u_b$ and return to
$u=0$ for a fixed $\alpha$.
For $u=0$ which is the initial and final
value of
$u$, different $\alpha$ correspond to same $J(p)$, so the different paths
characterized by different values of $\alpha$ have the same initial and
final points.
As we change $\alpha$ we change the path in
the complex $(u, \alpha)$ space keeping fixed the initial and the final
points. Since we do not cross any branch point
during such deformation of the path, we should expect to have the same
change in $\phi_c[p,J]$
for different $\alpha$.
As a result the final solution $\phi_0\chi(p)
+ h(p)$ should be independent of $\alpha$, even though
individually $\chi(p)$ and
$h(p)$ depend on $\alpha$.
In the absence of knowledge of the
$h(p)$ term, however, we cannot determine the explicit form of the
solution.

\subsection{Computation of  ratios of lump  tensions} \label{s3.4}

We shall now compute the tension associated with this
solution and compare with the known tension of the D-$(N_\parallel
-1)$-brane. For this we need to evaluate the effective action
$\Gamma[\phi]$ at
$\phi=\phi_0\chi(p)+h(p)$. Since at the solution of the classical
equations of
motion $\Gamma[\phi] = - W[J]$, the value of the action for this solution
is given by:
\be \label{e39}
\Gamma[\phi_0\chi(p)+ h(p)] = -\lim_{\uu\to 0'} W[\uu
\psi(p)]\, .
\ee
In taking the $\uu\to 0'$ limit, we need to choose the branch in the
complex $\uu$ plane along
which $F(\uu)$ approaches $\phi_0$. Now from eq.\refb{e7} we see that on
the trivial branch,
\be \label{e40}
W[\uu\psi(p)]
= \sum_{n=2}^\infty {\uu^n \over n!} \int
\Bigl( \prod_{k=1}^n d^Np_k \Bigr)\,
g^{(n)}(p_1, \ldots p_n) \psi(p_1)\ldots \psi(p_n) \delta(\sum_{i=1}^n
p_i) \, .
\ee
Using the form \refb{e24} for $g^{(n)}$ and \refb{e18} for $\psi(p)$, and
explicitly performing the integration over the $p_{i\parallel}$, we
get
\be \label{e41}
W[\uu\psi(p)] =
\delta_\parallel(0) \sum_{n=2}^\infty {1 \over
n!} C_n\uu^n K^n \int \prod_{k=1}^n d^{N_\perp} p_{k\perp} \,
\delta(\sum_{i=1}^n p_{i\perp})
\exp\Big({\sum_{k,l=1}^n a_{k-l} p_{k\perp} \cdot p_{l\perp}}\Big) \, .
\ee
$a_k$'s have been defined in eq.\refb{e26}, and $\delta_\parallel(0)$
stands for $\delta(p_\parallel=0)$. Using the standard finite volume
regularization we can identify $\delta_\parallel(0)$ with $V_\parallel /
(2\pi)^{N_\parallel}$, where $V_\parallel$ is the total volume spanned by
the coordinates $x_\parallel$. The integral appearing in eq.\refb{e41}
is the same
integral
that appears in eq.\refb{e27} with $x_\perp$ set to zero. Performing the
integral as before, we get
\be \label{e42}
W[\uu\psi(p)] =
{V_\parallel \over
(2\pi)^{N_\parallel}} \sum_{n=2}^\infty {1 \over
n!} C_n \uu^n  K^n \Big\{{1\over n}
\Big(\prod_{s=1}^{n-1}
{\pi\over b_s}\Big)\Big\}^{N_\perp/2}  \, .
\ee
We now use eqs.\refb{es2a}, \refb{es2}, \refb{es3} to write this as
\be \label{e43}
W[\uu\psi(p)] =
{V_\parallel \over
(2\pi)^{N_\parallel}} \sum_{n=2}^\infty {1 \over
n!} C_n \uu^n \exp\Big[{N_\perp\over 2} \big(B(\alpha) +
s_n(\alpha)\big)\Big]\, ,
\ee
where
the coefficients $s_n$, introduced in \refb{es2}
approach zero as $n\to\infty$.
If we define
\be \label{e44}
G(\uu) = \sum_{n=2}^\infty {1 \over
n!} C_n \uu^n \, ,
\ee
and
\be \label{e44a}
P(\uu) = \exp\Big( {N_\perp\over 2} B(\alpha) \Big)
\sum_{n=2}^\infty {1 \over
n!} C_n \uu^n \Big(e^{N_\perp s_n(\alpha)/2} - 1\Big) \, ,
\ee
then we can rewrite \refb{e43} as:
\be \label{e43a}
W[\uu\psi(p)]
 =
{V_\parallel \over
(2\pi)^{N_\parallel}} \Big[\exp\Big( {N_\perp\over 2} B(\alpha) \Big)
G(\uu) + P(\uu)\Big]\, .
\ee
Since $\lim_{n\to\infty} s_n=0$,
the coefficients of the Taylor
series expansion in $\uu$ of $P(\uu)$
grows at a rate slower than
that of $G(\uu)$.

Comparison with eq.\refb{e15} shows that $F(\uu) = dG(\uu) / d\uu$.
Thus $G(\uu)$ must also have a branch point at $\uu=\uu_b$. On the
other hand, since the Taylor series expansion of $P(\uu)$ converges
faster than that of $G(\uu)$, $P(\uu)$ has no branch point for
$|\uu|<|\uu_b|$, and may or may not have a branch point at
$\uu=\uu_b$. We now analytically continue $\uu$ around $\uu_b$
and
return to the origin. Suppose during this process $G(\uu)$ returns to
some constant $G_0$. It is important to note that $G_0$
does not depend on $N_\perp$, as is clear from eqn.~\refb{e44}.
If $P(\uu)$
does not have a branch point at
$\uu_b$ then
it returns to zero, otherwise it returns to some value $P_0$.
Thus using
\refb{e39} and \refb{e43a} we get
\be \label{e45}
\Gamma[\phi_0\chi(p)+ h(p)] = - {V_\parallel \over
(2\pi)^{N_\parallel}} \Big( e^{N_\perp B(\alpha) / 2}
G_0 + P_0\Big) \, .
\ee
We shall
refer to $P_0$ as
the sub-leading
contribution since it comes from the function $P(\uu)$
with a milder singularity at $\uu_b$ than $G(\uu)$.
Note that $P(u)$ and hence $P_0$ vanishes
for $N_\perp=0$.
Thus if
$P_0$ does not vanish it is necessarily $N_\perp$ dependent.
We shall now make the final assumption in our analysis, namely that {\em
the sub-leading
contribution $P_0$ vanishes}.\footnote{This
is not in contradiction with the earlier claim (previous subsection) that
the sub-leading contribution $h(q)$ to the classical solution
must be non-vanishing so that
$\phi_0\chi(q) + h(q)$ is independent of $\alpha$. Indeed,
$P_0$ receives contribution only from the sub-leading terms $s_n$ in
$f(\alpha, n)$, whereas
$h(q)$ receives contributions also from the sub-leading terms
in $k(\alpha, n)$.}
Using this critical 
assumption in
eq.\refb{e45},
we can identify the tension of the 
solitonic $(N_\parallel-1)$ brane to
be:
\be \label{e46}
\TT_{N_\parallel -1} = {1 \over
(2\pi)^{N_\parallel}} \, G_0 \, e^{(N
- N_\parallel) B(\alpha) / 2}
\, ,
\ee
where we have used  $N=N_\parallel+N_\perp$.
We now note that the function
$G(\uu)$ and hence the constant $G_0$ does not depend on $N_\perp$.
Thus eq.\refb{e46} gives the prediction:
\be \label{e47}
{1\over 2\pi} {\TT_{N_\parallel-2}\over \TT_{N_\parallel -1}}  =
e^{B(\alpha) / 2}\,.
\ee
Using the numerical result that $B(\alpha)$ vanishes identically, we see
that the right hand side of eq.\refb{e47} becomes 1.
This is in perfect agreement with the exact answer.

With our present knowledge, however, we cannot prove that $P_0$
vanishes.
Nevertheless, the remarkable agreement described above can be taken
as an
evidence for the underlying assumptions in our analysis, namely that the
quadratic term in the string field theory action expanded around the
tachyon vacuum is made solely of the ghost operators, and also that the
non-leading contribution from the function $P(u)$ vanishes at
$u=0$.
Further analysis of the non-leading terms $s_n$ is required
to establish if $P_0$ vanishes. In fact, even the knowledge of
$C_n$ may be important for this analysis, in which case a specific
choice of
the kinetic operator would be required.

\sectiono{Further studies and open questions} \label{s4}

In this section we discuss important issues for which
our understanding is fairly incomplete. We begin by exploring
how much freedom there is in choosing a kinetic operator
built
exclusively from the ghost sector. Then we discuss
some attempts to use level expansion to understand the
proposed actions around the tachyon vacuum. We discuss some
difficulties with the use of the Siegel gauge, and some
alternative gauges. Finally we give some remarks on extensions
to superstring theory, and on implications for
the search of closed strings states.

\subsection{On the uniqueness of $\QQ$} \label{s4.1}

Our discussion of  kinetic operators $\QQ$ that are
constructed purely of ghost operators identified a family
of them, and thus the general form in \refb{choice}
$\QQ = \sum_{n\geq 0} a_n \CC_n$. More general $\QQ$'s made
of purely ghost operators might exist, but we will focus here
on this family.

The key question is whether all these operators are
actually equivalent. If they are, anyone of them could
be chosen, making $\QQ = c_0$ the most obvious choice.
If they are not equivalent, one must search for the
correct one.

As discussed in section \ref{s2.2} and in particular
in equations \refb{efderp},
field redefinitions preserving
the cubic structure of the theory arise from BPZ odd ghost
number zero operators
that are derivations of the star algebra.  There is a well-known
family of such operators, they are conformal
transformations
that leave the string midpoint fixed~\cite{Witten:1986qs}
\be
\label{kns}
 K_n \equiv L_n - (-1)^n L_{-n}\, , \qquad  n \geq 1\,.
\ee
With $\CC_n = c_{n} + (-)^n c_{-n}$ we have the algebra
\be
\label{alg}
[K_n\, , \CC_m \,] = -[ (2n + m)\, \CC_{m+n} + (-)^n  (2n - m)\,
\CC_{m-n}\,]\,.
\ee
For example $[K_n, \CC_0] = - 4 n \,\CC_n$, and this indicates
that the operator $\QQ = \CC_0$ can be deformed infinitesimally
in every direction except along itself by a field
redefinition of the
form $\Psi\to
\exp(\epsilon_n K_n)\Psi$. A change of $\QQ$ along
itself corresponds to a change in coupling, and we do not
expect such changes to be possible by field redefinitions.

For more general $\QQ$,
even the infinitesimal deformation problem is
somewhat
nontrivial. Consider an arbitrary $\QQ = \sum_{n\geq 0} a_n \CC_n$
and an arbitrary deformation into $\QQ' = \QQ + \epsilon\Delta \QQ$ with
$\Delta \QQ = \sum_{n\geq 0} e_n \CC_n$,
where only a finite number of
$e_n$'s are nonzero. We ask if
$\QQ$ and $\QQ'$ are equivalent up to an infinitesimal scaling. With $K=
\sum_n d_n K_n$ such equivalence will hold if there
exist constants $d_n$'s
and a constant $r$ such that
\be
e^{\epsilon K} \QQ  e^{-\epsilon K} = (1+\epsilon \,r) \QQ' \quad
\to \quad [K, \QQ] = r \QQ  + \Delta \QQ \,.
\ee
If the solution involves an infinite number of non-vanishing
$d_n$'s, one must study their large $n$ behaviour to decide if
the infinitesimal
conformal transformation associated to $K$ exists. While it
follows from  the earlier remarks that for
$\QQ = c_0$  all infinitesimal deformations
with $e_0=0$
yield equivalent
kinetic operators,
we suspect that this will not be the
case for general $\QQ$.
One way
to see this is to recall that the identity string field
$ \langle \II |$ is not annihilated by
$c_0$ \cite{0006240}:
\be \label{econv1}
\langle \II |  \, c_0  \not= 0\,.
\ee
This
is some sort of anomaly, since any derivation of the star product,
such as $c_0$,
would be expected to annihilate the identity of the
product. On the other hand (\cite{0006240},
eqn. (6.16))
\be  \label{conv}
\langle  \II    | \;\Bigl(c_0 + {1\over 2} (c_2 + c_{-2} )
\Bigr) \;=0\,.
\ee
Since  $\langle \II | K_n = 0$ for all $n$, we cannot expect that
\be
(\CC_0 + \CC_2 + \epsilon\, \CC_0) \sim \exp (\sum d_n K_n)\,
(\CC_0 +
\CC_2)\, \exp (-\sum d_n K_n)\,
\ee
with the exponentials defining regular transformations. This is clear
since the right hand side would annihilate the identity while the
left hand side would not.

It would therefore appear that up to rescaling
there are at least two
classes
of inequivalent kinetic terms, those which annihilate the identity
and those which do not. In fact, ref.~\cite{HORO}
restricts its
attention to those operators that annihilate the identity, as this
string field plays an important role in the conjectured purely
cubic form of SFT.

\subsection{On the use of level truncation} \label{s4.2}

For the standard cubic open string field theory with
the usual BRST operator $Q_B$, there is overwhelming
numerical evidence (but yet no proof) that the level
truncation scheme \cite{KS}
provides results that
converge  to the correct non-perturbative answer.
It is therefore of  interest to ask whether level truncation
can be used  to test possible choices for the BRST operator
$\QQ$ around the tachyon vacuum.
So far we have only obtained some
partial and somewhat problematic
results for the simplest choice
$\QQ = \tilde A \, c_0$. We
allow for the presence of an unknown
normalization constant $\tilde A$.
By a rescaling of the string field, we
choose to write $\QQ = c_0$ and put an overall costant
$A=\wt A^3$
in front of the string field theory action.

We first look for a translationally invariant solution
of the truncated string field
theory which corresponds to the space filling D-25 brane --
the
original
vacuum before condensation. The potential for a static
configuration is
\be \label{cubicc0}
-S(\Psi) =
\frac{1}{g_0^2 \,2 \pi^2} \, {\cal V} (\Psi)
\ee
with
\be
{\cal V}(\Psi) = A \cdot 2 \pi^2 \left[
\frac{1}{2} \langle \Psi, c_0 \Psi \rangle
+ \frac{1}{3} \langle \Psi, \Psi * \Psi \rangle    \right]\,.
\ee
By construction, the zero of the energy
 corresponds to the tachyon vacuum $\Psi =0$, ${\cal V }(\Psi=0) = 0$
and all physical excitations are expected to have
positive energy above this ground state.
 We adopt the normalization conventions of \cite{9911116},
in which the space filling solution $\Psi_{D_{25}}$ obeys
${\cal V}(\Psi_{D_{25}}) = 1$. This in turn should
allow to fix the constant $A$.

We work in the Siegel gauge $b_0 \Psi =0$.
In the level $(0,0)$ approximation, the effect of replacing
$Q_B$ by $c_0$ is simply to change the sign of the kinetic
term. From the results in \cite{9912249}
we readily find the translationally invariant solution
$\Psi^{(0,0)} = t_c \, c_1 |0 \rangle$ with $t_c =-(4 \sqrt{3}/9)^3$.
This gives ${\cal V}^{(0,0)} \approx 0.684\, A$.
We simply quote the results that we find at higher levels:
 ${\cal V}^{(2,6)} \approx 0.1849\, A$,
 ${\cal V}^{(4,8)} \approx 0.09656\, A$.
Unlike the case of the action with the standard BRST operator $Q_B$,
here we do not find any evidence of convergence to a finite limit
in the level truncation. Rather, the energy of the
space filling brane seems to be pushed towards zero as the level
is increased.

We have also looked for solitonic solutions $\Psi_{D_{24}}$
corresponding to D-24 branes,
in which the string field has the shape of a lump along a  direction
$X$ and approaches asymptotically the tachyon vacuum.
As in \cite{0005036}, in order to implement
systematically the level truncation
scheme we compactify the direction $X$
on a circle of radius $R$. For each
zero momentum state $|\Psi_i \rangle$ that we had before,
we obtain a tower of Fourier modes $|\Psi_{i,n} \rangle$,
$n=0,1,2 \dots$, of momentum $p = n/R$ along the $X$ direction.
Level is now defined as $n^2/R^2 + N_i -N_0$,
where $N_i$ is the number operator and $N_0=-1$ is the
number eigenvalue
of the zero momentum tachyon.

We have considered
the case $R = \sqrt{3}$ and performed the computation
up to level $(3 ,6)$.
As for finite $A$ the energy of the D25-brane
appears
to go to zero in our calculations, we considered an energy
ratio where this fact would not affect the result.
Indeed, the ratio
\be
{\cal R}^{(N,M)} = \frac{R\, {\cal V}(\Psi^{(N,M)}_{D_{24}})}{
 {\cal V}(\Psi^{(N,M)}_{D_{25}})} \, ,
\ee
does not depend on the unknown
constant $A$ and is expected to converge to 1. We find:
${\cal R}^{(1/3, 2/3)} \approx 1.11056$, ${\cal R}^{(4/3,8/3)}
  \approx 0.958355$,
${\cal R}^{(7/3, 14/3)} \approx 1.07661$,
${\cal R}^{(3,6)} \approx 1.2141$.
Although not far from the expected answer,
it is not clear we are finding convergence to
the correct value.

The conclusion is
that level truncation
with  $\QQ =c_0$ and in the Siegel gauge is
somewhat problematic.
This could be (i) a general problem with level expansion
around the tachyon vacuum whenever the kinetic term
has been brought to a purely ghost form, (ii) a problem
with the specific choice $\QQ= c_0$ as opposed to other
choices in the general family $\QQ = \sum a_n \CC_n$,  or, (iii) a
problem with the Siegel gauge. Indeed,
there are reasons to believe
that this may not be a legal choice, as we shall discuss
in the next subsection.

\subsection{On gauge fixing} \label{s4.3}

Our analysis of the tachyon effective action in section 3
assumes that given a string field theory with a purely
ghost BRST operator $\QQ$,
one can find a suitable gauge fixing
which does not involve matter operators. We believe
on general grounds that this should always be possible.
However we would like to point out some complications
which arise in the application of the standard Siegel gauge
to string field theories with non-standard BRST operators,
like the family $\QQ = \sum_{n \geq 0} a_n  \CC_n $.

We focus again on the simplest case of the string field theory
with $\QQ =c_0$. Let us consider imposing
the Siegel gauge condition $b_0 \Psi= 0$.
This is always possible
at the linearized level, since
if $b_0 \Psi \not=0$, then
$\Psi_S = \Psi - \QQ (b_0\Psi)=  \Psi - c_0 b_0 \Psi$ satisfies
$b_0 \Psi_S = 0$. Moreover, the Siegel condition
fixes the gauge completely.
Indeed if $\delta \Psi$ is pure gauge, and in Siegel gauge, then
$\QQ \, \delta \Psi = 0$ and $b_0 \, \delta \Psi =0$, leading
necessarily to  $\delta \Psi =0$.\footnote{
As long as $L_0 \Psi \neq 0$,
this analysis is entirely parallel
to the one with $Q_B$, for which $\{ b_0,  Q_B \} = L_0$. }

Consider now computing perturbative amplitudes in this string
field theory.
We shall find that
tree level Green's functions
diverge in the Siegel gauge.
The only modification of the conventional Feynman rules
\cite{WITTENBSFT}
is that the propagator is $b_0$ instead of $b_0 L_0^{-1}$.
This is however a dramatic difference. The factor $L_0^{-1} =
\int_{0}^{\infty} e^{-t L_0}$ had the interpretation
of  an integral over strips of
any length. In the present case,
the strip has collapsed to zero length, and
there is no integration over moduli space, but simply a ghost
insertion $b_0 = \int d\sigma ( b_{zz} + b_{\bar z\bar z})$
across the collapsed strip.  Any term in the
$n$-point amplitude
requires $(n-3)$
ghost insertions of the form
\be
\int_{\gamma/2} \Bigl( dz\; b(f_k^{(n)}(z))\; z \, \Bigl(
\frac{df_k^{(n)}}{dz} \Bigr)^2 \, +\, {a.h.} \Bigr)
\ee
where $f_k^{(n)}(z)$ were defined in (\ref{e20}).
Each integral across a collapsed strip  has
been referred to the local
coordinates $z$ of the relevant half-strings. One of these contributions is
what is shown in the above formula. Thus
$\gamma/2$ means either the right or
the left half-boundary of a
canonical upper-half unit disk. In addition, the
$b(z)$ appearing in this half-string integral has been mapped into the
global uniformizing disk where all the strings are glued together.
This is
necessary because all  correlators are computed on the uniformizing
disk.
Near $z=i$ the integrand behaves as  $\sim  (1+iz)^{-2+4/n}$,
which is a non-integrable singularity for $n \geq 4$. Thus
in Siegel gauge, the coefficients $C_n$ in (\ref{e19}) appear to
diverge.

There is a simple cure for this problem, which consists
of considering a gauge-fixing condition $B \, \Psi =0$
with $B = \oint z g(z)  b(z)$, where $g(z) \sim (1+iz)^m$
near $z \sim i$, and $m >1$. We also demand
that $g(z)$ contains a constant in its Laurent
expansion, so that $B$ has some amount of $b_0$:
this is required to have $\{ B, c_0 \} \neq 0$
so that the gauge condition can be imposed (at least
at the linearized level). For example, with
$g(z) = 1 + (z^2 +1/z^2)/2$ we have $B=b_0 + (b_2 +b_{-2})/2$.
The extra double zero $\sim (1+iz)^2$ of the propagator
near $z=i$ ensures now the convergence of string amplitudes.
It would be interesting to explore these `finite' gauge
choices in the level truncation scheme.

\subsection{On superstring field theory and closed strings}
\label{s4.4}

\bigskip
As was the case for the bosonic theory we may conjecture that at the
tachyon vacuum the kinetic term of superstring field theory does not
involve a matter sector operator.
We will offer some remarks that may help in constructing
an appropriate action.

The algebraic
structure of the open Neveu-Schwarz (NS) superstring field theory of
refs.~\cite{9503099,9912120,9912121} demands the existence of two
Grassmann odd operators
$Q_B$ and $\eta_0$ anticommuting with
each other, both of which square to zero, satisfy the same BPZ
conjugation property, and are derivations of the star algebra. The
operator
$\eta_0$, having picture number minus one, is the zero mode of the
Grassmann odd  field $\eta(z)$ arising from fermionization of the
superghosts. If the BRST operator $Q_B$ is replaced by $c_0$ or any of the
$\CC_n$ operators, the algebraic structure will remain in place, for this
operator would still commute with $\eta_0$ and, as we have discussed
before, satisfies all the other requisite properties.  Thus an action with
$\QQ=\sum_n a_n \CC_n$ would satisfy the conditions of gauge invariance.
Moreover, there would be no physical states around this vacuum.  Indeed,
in this string field theory the linearized equations of motion
are  $\QQ \eta_0 \Phi =0$ where $\Phi$ is a string field of ghost number
zero and picture number zero in the so-called large Hilbert space.
The gauge invariance $\delta \Phi = \eta_0 \Omega$ allows one to
write $\Phi = \xi_0 V$ where $V$ is a conventional picture number
minus one field of the NS sector. In this partial gauge the equation
of motion reduces to $\QQ V=0$ which, for $\QQ= Q_B$  is the standard BRST
cohomology problem in the conventional small Hilbert space. If $\QQ=
\sum a_n \CC_n$, we will have no physical states.

While it
seems tempting to simply replace $Q_B$ by the appropriate ghost
operator to obtain an action around the tachyon vacuum,
there are some  questions that suggest that this change
may not suffice to obtain the desired action. In particular,
the Chan-Paton like factors that implement the
$Z_2$ symmetry under which the GSO odd states change
sign \cite{0001084,0002211} must somehow change in the tachyon
vacuum. Indeed, while the tachyon potential is even with respect to
the field variable which vanishes
at the unstable vacuum, it is not
even with respect to the field variable which vanishes
at any one of
the tachyon vacua.

It is perhaps worthwhile to note that the replacement
$Q_B \to \sum a_n \CC_n$ does not preserve the algebraic
structure of cubic superstring field theory \cite{Witten:1986qs}.
The problem is that the picture changing operator $X(z)$, that must
be inserted at the string midpoint in the definition of the
star product, while BRST invariant, is
not $\CC_n$ invariant.
This spoils the derivation property which is necessary for gauge
invariance.  This cubic string field theory, in any case, has long been
known to be  problematic \cite{wendt}, and does not appear to give a
sensible tachyon potential either \cite{0004112}. Interestingly,
the double picture changing insertion used in the cubic superstring field
theory action advocated in refs.\cite{AREF1,AREF2,0011117} does commute
with the $\CC_n$ operators, and it therefore appears that gauge
invariant
actions without cohomology could be constructed.

\medskip
We can now ask if a similar replacement of the BRST operator with
another having trivial cohomology can be done in closed string field
theory \cite{9206084}. The answer is no. The problem here is that
the  interactions beyond the cubic one involve the integration of
correlators over finite pieces of the moduli spaces of Riemann surfaces.
Key to the algebraic structure of the theory is the fact that the BRST
operator acts as a total derivative on moduli space.  This arises because
forms on moduli space involve the antighost field $b(z)$ and
$\{ Q_B,b(z) \} = T (z)$, where the energy momentum tensor $T$
generates deformations of correlators. Such property will not
hold upon replacing $Q_B$ with an object constructed purely from
ghosts.

\medskip
A final question is whether closed strings arise from the kind
of open string field theory we have been considering here, namely
one with a kinetic opertor having no cohomology. We have no
definite answer. Recall that in cubic open string field theory
closed string poles definitely arise in string
loop amplitudes~\cite{Freedman:1988fr}.
A subset of open string propagators in a string diagram,
represented by
a strip of total length $T\in [0, \infty]$, whose middle line  defines a
nontrivial closed curve, will give rise to closed string poles via
integration over the neighborhood of $T=0$. In the kind of actions we
have considered here,  the propagator
collapses to an overlap with some antighost insertion. It does not seem
possible to obtain closed string poles from an amplitude that just
includes the contribution from the point $T=0$. On the other hand, there
are well known complications with the precise Batalin-Vilkovisky
quantization of conventional cubic open string field theory, and they
precisely arise from the $T\to 0$ limit of a one-point one-loop amplitude
\cite{thorn,9705241}. The whole issue surrounding closed strings in open
string field theory after tachyon condensation is subtle enough
\cite{0012081,9901159,0002223,0009061,0010240}
that  further work seems necessary to attain definite conclusions.

\subsection{Concluding remarks}

The open string field theories studied in this paper
manifestly implement the absence of conventional open
string dynamics on the vacuum of the tachyon. They naturally
implement gauge invariance and are rather reminiscent of
the $p$-adic string theory, where after a simple invertible
field redefinition the kinetic operator is just a constant.
In addition,
they are in many ways strikingly simple, much more so than
the conventional cubic open string field theory in the background of a
D-brane. For example, while we were able here to compute the
exact momentum dependence of the off-shell $n$-tachyon amplitudes,
such computation appears to be prohibitively complicated in conventional
cubic string field theory.  If the normalization of the
$n$-tachyon amplitudes can be computed, perhaps along the lines
of comments in sect.~\ref{s4.3},  we may then be able to
find the explicit solutions representing D-branes, to obtain the precise form
of the kinetic term $\QQ$,  and to settle the outstanding assumptions of our
analysis.

\medskip
\bigskip
\noindent {\bf Acknowledgements}: We would like to thank D.
Ghoshal, D. Jatkar,  N.~Moeller,  P. Mukhopadhyay, S. Panda, and
W.~Taylor for
useful discussions. The work of L.R. was supported in part
by Princeton University
``Dicke Fellowship'' and by NSF grant 9802484.
The work of  B.Z. was supported in part
by DOE contract \#DE-FC02-94ER40818.

\end{document}